\newcounter{myctr}

\documentclass{ws-acs}
\usepackage{url}

\begin{document}

\makeatletter
\def\@biblabel#1{[#1]}
\makeatother

\markboth{RC Ball, M Diakonova, RS MacKay}{Quantifying Emergence in terms of Persistent Mutual Information}

%
\catchline{}{}{}{}{}
%

\title{QUANTIFYING EMERGENCE IN TERMS OF PERSISTENT MUTUAL INFORMATION}

\author{\footnotesize RC BALL}
\address{Centre for Complexity Science and Department of Physics, University of Warwick\\
Coventry, CV4 7AL,
UK \\
R.C.Ball@warwick.ac.uk}

\author{M Diakonova}
\address{Centre for Complexity Science and Department of Physics, University of Warwick\\
Coventry, CV4 7AL,
UK \\
M.Diakonova@warwick.ac.uk}

\author{RS MacKay}
\address{Centre for Complexity Science and Department of Mathematics, University of Warwick\\
Coventry, CV4 7AL,
UK \\
R.S.MacKay@warwick.ac.uk}

\maketitle

\begin{history}
\received{(received date)}
\revised{(revised date)}
\end{history}

\begin{abstract}
We define Persistent Mutual Information (PMI) as the Mutual (Shannon)
Information between the past history of a system and its evolution
significantly later in the future. This quantifies how much past observations
enable long term prediction, which we propose as the primary signature
of (Strong) Emergent Behaviour. 
The key feature of our definition
of PMI is the omission of an interval of 'present' time, so that the mutual information between close times is excluded:
this renders PMI robust to superposed noise or chaotic behaviour or graininess of data, distinguishing it from a range
of established Complexity Measures. For the logistic map we compare predicted with measured long-time PMI data. We show that measured PMI data captures not just the period
doubling cascade but also the associated cascade of banded chaos, without confusion by the overlayer
of chaotic decoration. We find that the standard map has apparently infinite PMI, but with well defined fractal scaling which we can interpret in terms of the relative information codimension.  Whilst our main focus is in terms of PMI over
time, we can also apply the idea to PMI across space in spatially-extended
systems as a generalisation of the notion of ordered phases.
\end{abstract}


\section{Introduction}
Our starting point is the desire to discover and quantify the extent
to which the future evolution of a dynamical system can be predicted
from its past, and from the stand point of Complexity Theory we are
interested in assessing this from observed data alone without any
prior parametric model. 

We should nevertheless admit prior classification as to the general
nature of the system, and simple constraining parameters such as its
size, composition and local laws of motion. Given that information,
there may prove to be entirely reproducible features of its subsequent
evolution which inevitably emerge over time, such as eventual steady
state behaviour (including probability distributions). This we follow
\cite{Chalmers,BY} and others in terming \emph{weak emergence}. The emergence is weak
in terms of there being no choice of outcome, it can be anticipated
without detailed inspection of the particular instance.

We focus on \emph{Strong Emergence} by which we mean features of behaviour
significantly into the future which can only be predicted with knowledge of
prior history. The implication is that the system has made conserved
choices not determined by obvious conservation laws, or at least
nearly conserved choices which imply the existence of associated slow
variables.

More formally, we must conceive of an ensemble comprising a probability distribution
of realisations of the system (and its history), from which observed
histories are drawn independently. The behaviour of a particular realisation
which can be anticipated from observing other realisations is weakly
emergent, whilst that which can only be forecast from the observation
of the past of each particular instance is strong emergence. A related distinction between weak and strong emergence is given in \cite{M08}, but with quantification based on a metric on the underlying space, rather than purely measure-theoretic.

\section{Persistent Mutual Information}
Within an ensemble of histories of the system we can quantify strong
emergence in terms of mutual information between past and future history
which persists across an interval of time $\tau$. This \emph{Persistent
Mutual Information} is given by

\begin{equation}
I(\tau)=\int \log \left(\frac{P[x_{-0},x_{\tau+}]}{P[x_{-0}]P[x_{\tau+}]}\right)P[x_{-0},x_{\tau+}]dx_{-0}dx_{\tau+}  \label{PMI}
\end{equation}
where $x_{-0}$ designates a history of the system from far past up
to present time $0$, $x_{\tau+}$ is the corresponding history of
the system from later time $\tau$ onwards, $P[x_{-0},x_{\tau+}]$
is their joint probability density within the ensemble of histories,
and $P[x_{-0}]P[x_{\tau+}]$ is the product of corresponding marginal
probability densities for past and future taken separately. If the
history variables $x(t)$ are discrete-valued then integration over
histories is interpreted as summation; in the continuous case $I(\tau)$
has the merit of being independent of continuous changes of variable,
so long as they preserve time labelling.

Quantitatively $I(\tau)$ measures the deficit of Shannon Entropy
in the joint history compared to that of past and future taken independently,
that is 

\begin{equation}
I(\tau)=H\left[P[x_{-0}]\right]+H\left[P[x_{\tau+}]\right]-H\left[P[x_{-0},x_{\tau+}]\right] \label{PMIEntropy}
\end{equation}
where the separate Shannon Entropies for a probability density $P$ of a set of variables $y$ are given generically by 
\begin{equation}
H\left[P\right]=-\int \log\left(P[y]\right)P[y]dy . 
\end{equation}
Thus it is precisely the amount of information (in Shannon Entropy)
about the future which is determined by the past, and hence \emph{the
extent to which the future can be forecast from past observations}
(of the same realisation).

The key distinguishing feature of our definition above is the exclusion
of information on $x_{0\tau}$, that is the intervening time
interval of length $\tau$. This ensures that $I(\tau)$ is only sensitive
to system memory which persists right across time $\tau$; any
features of shorter term correlation do not contribute. The choice
of $\tau$ must inevitably be informed by observation, but the extreme
cases have sharp interpretation.

 $I(0)$ corresponds directly to the Excess Entropy as introduced in  \cite{G1} where it was called ``Effective Measure Complexity'', and makes no distinction
of timescale in the transmission of information. This quantity has been discussed in many guises: as effective measure complexity in \cite{G1,LN}, ``predictive information'' in \cite{BN} and as Excess Entropy in  \cite{CP,FMC}, to name but a few. See also \cite{G3,G2,FMC} for measurements of Excess Entropy and the related Entropy Rate on a variety of systems, including the logistic map. 

Our sharpest measure of Strong Emergence is the \emph{Permanently
Persistent Mutual Information (PPMI)}, that is the PMI $I(\infty)$
which persists to infinite time.\emph{ }This quantifies the degree
of permanent choice spontaneously made by the system, which cannot
be anticipated without observation but which persists for all time.
A prominent class of example is spontaneous symmetry breaking by ordered
phases of matter: here a physical system is destined to order in a
state of lower symmetry than the probability distribution of initial
conditions, and hence must make a choice (such as direction of magnetisation)
which (on the scale of microscopic times) endures forever. As a result Strong Emergence can only be diagnosed by observing multiple independent realisations of the system, not just one long time history. An interesting though anomalous case is presented by clock phase, where time shift leads to different phases. This is exploited in measuring PPMI for the logistic map in the following section (see fig. \ref{fig1}). 

PPMI corresponds to some partitioning of the attracting dynamics of the system into negligibly communicating (and negligibly overlapping)
subdistributions. If the dynamics evolves into partition $i$ with
probability $p_{i}$ then the PPMI is simply given by 
\begin{equation}
I(\infty)=-\sum_{i}p_{i} \log \left(p_{i}\right)
\end{equation}
which is the entropy of the discrete distribution $p_{i}$.
For deterministic dynamics each $p_{i}$ is simply determined by the sampling
of its associated basin of attraction in the distribution of initial
conditions, so in this case the PPMI is sensitive to the latter. However
for stochastic dynamics it is possible for the $p_{i}$ to be predominantly
determined by the distribution of early time fluctuations.

\section{PPMI in the logistic map}

We consider time series from the logistic map $x_{n+1}=\lambda x_{n}(1-x_{n})$
as a simplest non-trivial example which brings out some non-trivial
points. Depending on the control parameter $\lambda$, its attracting
dynamics can be a limit cycle, fully chaotic, or the combined case
of banded chaos where there is a strictly periodic sequence of bands
visited but chaotic evolution within the bands \cite{CE}. 

In the case of a periodic attracting orbit with period $T$, the choice
of phase of the cycle is a permanent choice leading directly to positive PPMI. For this case the phase separation is in the time domain, so the attractor can be fully sampled by shifting the start time of observation. If we assume the latter is uniformly distributed over a range of time large enough compared to the period, then the observed phase is a result of a uniformly
selected and symmetry breaking choice in which each phase has $p_{i}=1/T$,
and this leads to PPMI

\begin{equation}
I(\infty)= \log \left(T\right) \,.   \label{PPMI}
\end{equation}
This is a generic result and not special to the logistic map. Because there is just an attracting orbit, the Excess Entropy gives the same value. 

For fully chaotic attracting dynamics such as at $\lambda=4$, we
have to be careful in principle about limits. Provided the probability
densities are measured with only limited resolution $\delta x$ in
$x$, then we expect past and future to appear effectively independent
for $\tau>\tau(\delta x)$ and hence $I(\tau)\rightarrow0$ and there
is zero PPMI. Thus for chaotic motion the associated values are quite
different: the Excess Entropy is positive and reflects the complexity
of its dynamics, whereas the PPMI is zero reflecting the absence of
long time correlations.

	\begin{figure}[th] 
	\centerline{\psfig{file=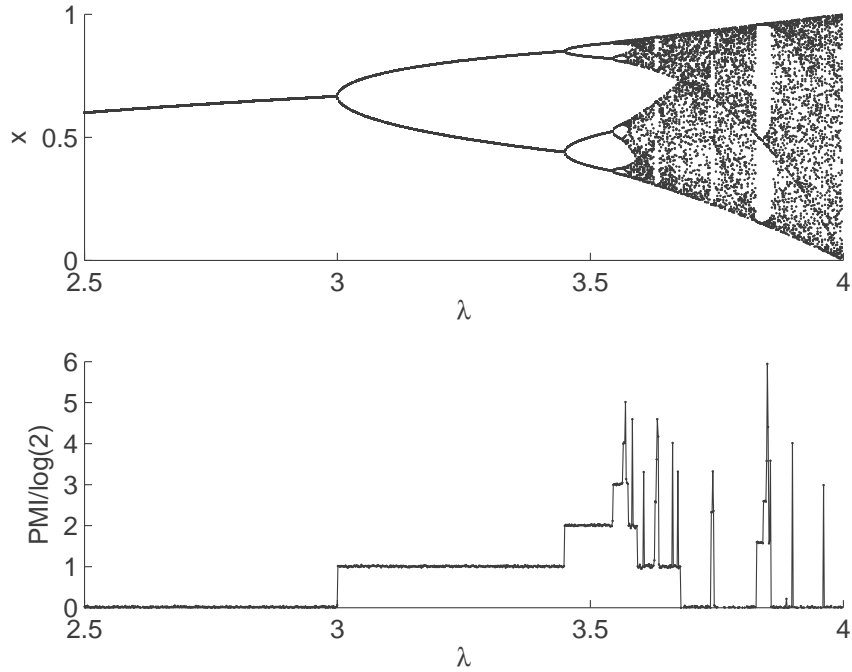,width=13cm}}
	\vspace*{8pt}
	\caption{Bifurcation diagram (above) and measured Persistent Mutual Information (below)  for the Logistic Map, as a function of the map control parameter $\lambda$.  For each value of $\lambda$ the map was allowed a minimum of $10^5$ iterations to settle and then the Mutual Information measured across a 'time' interval of $10^5$ iterations. Each MI measurement used the distance to 4th nearest neighbour to estimate probability density ($k=4$), based on a sample of $N=5000$ iterate pairs.  Before chaos sets in PMI increases stepwise in jumps of $\log{2}$, reflecting the doubling of the resolved period. It is also seen to pick up the band periodicity after the onset of chaos (resolving some more fine periods within the band hopping), as well as a nonchaotic period three regime.} \label{fig1}
	\end{figure}

\begin{figure}[th] 
	\centerline{\psfig{file=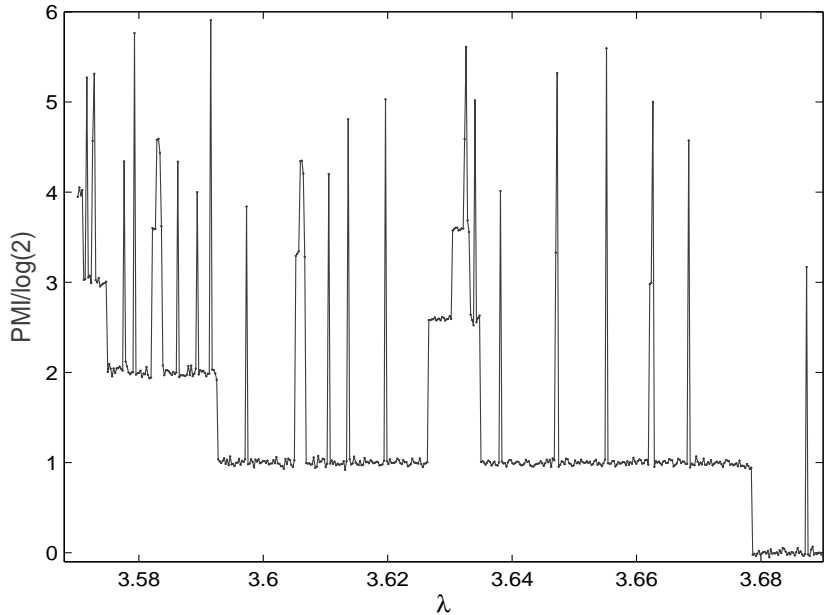,width=13cm, height = 9cm}}
	\vspace*{8pt}
	\caption{PMI for a chaotic region of the control parameter (sample size $N = 5000$, $k = 4$, settle time $10^{10}$, time separation $10^{6}$). In this range chaotic bands are known to merge (see bifurcation diagram in fig. \ref{fig1}. PMI picks out the relevant decrease in overall periodicity as 16 bands merge pairwise into eight, four, and finally two at $\lambda$ slightly less than $3.68$. PMI also detects a period tripling regime, which can be seen around $\lambda = 3.63$.}\label{fig2}
	\end{figure}

Both of the above results can be seen in measured numerical data for
the logistic map in fig. \ref{fig1}. What is more pleasing still is the
behaviour of PMI for banded chaos, where a T-periodic sequence of
bands shows through to give $I(\infty) \ge \log \left(T\right)$ (assuming
random initiation phase as before) with equality when the Tth iterate restricted to one band is mixing. The fact that the numerical results overshoot $ \log \left(T\right)$ for many parameter values can be attributed to the presence of a finer partition than the T bands, for example around an attracting periodic orbit of period a multiple of T or into sub-bands with a period a multiple of T. Even in cases where the dynamics really is T-periodic mixing (meaning it cyclically permutes T bands and the Tth iterate restricted to one is mixing), and hence PPMI is precisely $ \log \left(T\right)$, the numerics might pick up some long-time correlation that does not decay until after the computational time interval. One can see this in the more detailed data of fig. \ref{fig2} where there are background plateaux corresponding to the periodicity of the observed major bands, decorated by narrower peaks corresponding to higher periodicities. The steps in PMI are particularly clear where bands merge because these special points have strong mixing dynamics within each band cycle \cite{Mis}. Figures 1 and 2 for the PPMI can be usefully contrasted with the Excess Entropy graphed in figure 1 of \cite{FMC}. On the period-doubling side they display the same values (log of the Period), whereas in the regions of banded chaos the PPMI picks out the number of bands whilst the Excess Entropy is complicated by its sensitivity to short time correlations in the chaotic decoration.  

\section{Issues measuring PMI}

Measuring Mutual Information and in particular the implied measurement
of the entropy of the joint distribution suffers from standard challenges
in measuring the entropy of high dimensional data. The naive `histogram'
method, in which probability densities are estimated directly from
frequency counts in pre-selected (multidimensional) intervals is
easy to apply but can require very large sample sizes in order to
ensure that the significant frequencies are estimated from multiple
(rather than single) counts. In practice we found the $k$'th neighbour
approach of Kraskov et al \cite{Kraskov} a more effective tool (from now on referred to as the k-NN method). It is more limited in sample size due to unfavourable order of algorithm, but
this was outweighed by its automatic adjustment of spatial resolution
to the actual density of points.

The basis of the k-NN method is to estimate the entropy of a distribution from the following estimate of the logarithm of local probability density about each sampled point:

\begin{equation} 
\log(p) \simeq \log \left(\frac{k/N} {\epsilon_k^d}\right) + \left[\Psi(k) - \Psi(N) - \log \left( k/N \right)\right]
 = -\log \epsilon_k^d + \Psi(k) - \Psi(N) 
\end{equation}
where $N$ is the total number of sample points and $\epsilon_k^d$ is the volume of space out to the location of the $k$'th nearest neighbour of the sample point in question.  The combination $\frac{k/N}{\epsilon_k^d} $ in the first logarithm is simply interpretable as an amount of sampled probability in the neighbourhood divided by corresponding volume.  In the remaining term $\Psi(z)=\Gamma'(z)/\Gamma(z)\simeq \log(z) - \frac{1}{2z} \quad \rm{ as } \quad |z| \rightarrow \infty $ is the digamma function and the whole of this term is only significant for small $k$, where it corrects a slight bias associated with finite sampling of the neighbourhood \cite{Kraskov} .

We interpret $\log \left( N/k \right) \simeq  \Psi(N)-\Psi(k) $ as the (logarithmic) probability resolution in these measurements.  When $N$ and either $k$ is large or held fixed the variation of these two forms is equivalent and we generally show the first for simplicity of exposition.  The main exception is for entropy data on the standard map below, where taking data down to $k=1$ significantly enhances our appreciation of scaling and the use of the more accurately scaled second form is important.

Because PMI is invariant under changes of variable, there is considerable
scope for choice of how to parameterise past and future before feeding
into the PMI measurement. For the logistic map we exploited its deterministic nature, by which one value of $x_{n}$ provides all information
about the past up to iterate $n$ which is relevant to the future, and similarly all influence of the past on the future beyond iterate $n'$ is captured by the value of $x_{n'}$. Note however that we did not require to identify minimalist
causal states in the sense discussed in section (5) below.

For systems without known causal coordinates, the practical measurement
of PMI has a rich time parameterisation. In principle what we can
directly measure is the mutual information $I(t_{1},t_{2};t_{3},t_{4})$
between time intervals $[t_{1},t_{2}]$ and $[t_{3},t_{4}]$. If we
assume stationarity then this is more naturally parameterised as $I(\tau;T_{-},T_{+})$
in terms of the intervals $T_{-}=t_{2}-t_{1}$ and $T_{+}=t_{4}-t_{3}$
of past and future respectively as well as the intervening interval
$\tau=t_{3}-t_{2}$. Then the full PMI is defined as

\begin{equation}
I(\tau)=\underset{T_{-},T_{+}\rightarrow\infty}{\lim}I(\tau;T_{-},T_{+}) \,.   \label{PMI2}
\end{equation}
If the PPMI is desired it is computationally efficient to set $\tau\rightarrow\infty$ before
taking the limits above, because the dimension of space in which entropy
must be measured is set by $T_{-}+T_{+}$ alone. By contrast with PMI,  the ÔPredictive InformationÕ developed at length in reference \cite{BN} is in the present notation $I(0;T,T)$.

\begin{figure}[th] 
	\centerline{\psfig{file=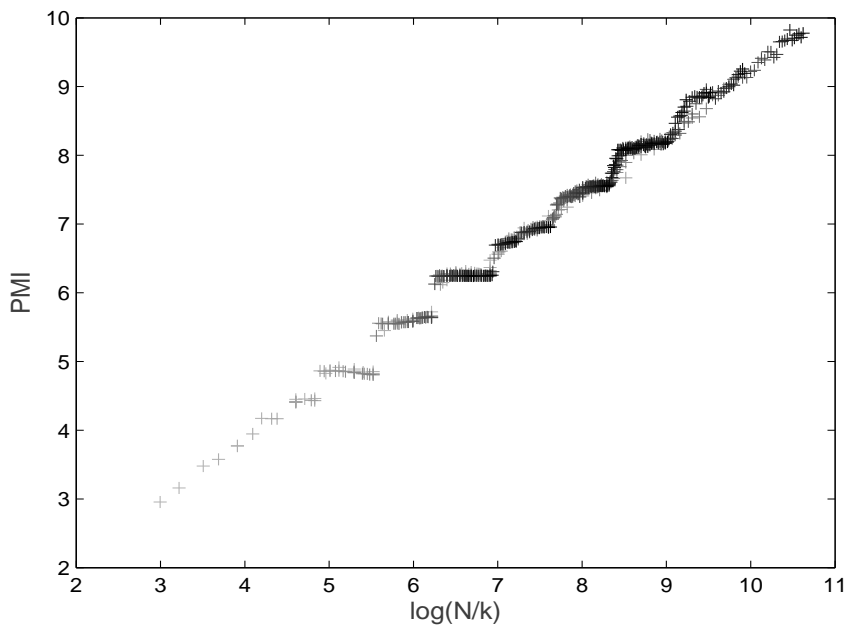,width=13cm, height = 9cm}}
	\vspace*{8pt}
	\caption{PMI of the logistic map at the period-doubling accumulation point plotted against the effective resolution with which probability density has been measured, given by $\log(N/k)$. ($N = 1000,2000, ..41000$ where runs with higher N correspond to darker points; $k = 1,2, 3, 4, 5, 10, .. 50$, time separation $10^{8}$, settle time $10^{10}$). The periodicity measured is resolution limited, with apparent overall slope of this plot ca $0.9$ in fair agreement with slope unity predicted on the basis of resolvable period $\propto N/k$ (see text).}    \label{figFP}
	\end{figure}

Practical measurement of PMI entails some limited resolution, whether
explicitly by histogram methods or implicitly through the depth of
sampling in k-NN and other adaptive methods. This inevitably
leads to long periodic orbits being capped in their apparent period
and hence their measured $I(\tau)$. We can be fairly concrete in
the case of measurement by the k-NN method, which looks
out across a neighbourhood whose aggregate measure is $k/N$. The
longest period one can thereby detect is of order $N/k$ so we are
led to expect $I(\tau)\simeq \log \left(\min\left(T,N/k\right)\right)$.
One point where we can check this quantitatively is the accumulation point of the period-doubling sequence \cite{CT,GR,F1}: fig. \ref{figFP} shows the measured results agreeing with the expectation that $I(\tau)\simeq \log \left(N/k\right)$.

\section{Relationship with Statistical Complexity}

Statistical Complexity (in certain contexts equivalent to ``True Measure Complexity'' first introduced in \cite{G1}; see also \cite{CY,shalizi}) is built on the projection
of past and future down to optimal causal states $S_{-}(t)[x_{t-}]$ such that $P[x_{t+}|x_{t-}]=P[x_{t+}|S_{-}(t)]$. In terms of these one readily obtains that the PMI is given
by the Mutual Information between time-separated forward and reverse
causal states, that is 

\begin{equation}
I(\tau)=H\left[P[S_{-}(t)]\right]+H\left[P[S_{+}(t+\tau)]\right]-H\left[P[S_{-}(t),S_{+}(t+\tau)]\right]  \label{I}
\end{equation}
as a straightforward generalisation of the corresponding result for
the Excess Entropy $I(0)$ \cite{G1}.

In general one cannot simplify the above formula ($I_{-+}$ in a natural
extension of the notation) to use other combinations of choice between
$S_{-}$ or $S_{+}$. However, we conjecture that under fairly general conditions they are all equivalent for the PPMI, i.e.  $\tau\rightarrow\infty$. It is an interesting open question whether for general time gap $\tau$ it can be proved that $I_{-+}\geq(I_{--},I_{++})\geq I_{+-}$, and perhaps also $I_{--}=I_{++}$. For $\tau =0$ similar forward and reverse time dependencies for $\epsilon$-machines have been considered in \cite{EMC}, where it was noted that in general $I_{--} \neq I_{++}$, and bidirectional machines were defined that incorporate this time asymmetry. 

\section{Fractal and Multifractal PMI: example of the standard map}

As we already observed for the accumulation points of the standard map,
where the attractor of the dynamics is a Cantor fractal adding machine, the measured
PMI may go to infinity as the resolution increases. The archetypal case
of this is where the probability measures themselves have fractal
support or more generally exhibit multifractal scaling. The general
phenomenology is that dividing space of dimension d into cells $c$ of linear width
$\epsilon$ of integrated measure $\mu_{c}=\int_{c}P(x)d^{d}x$, the density in a cell is estimated as $\mu_{c}/ \epsilon^{d}$ and hence to leading order as $\epsilon \rightarrow 0^{+}$
one expects
\begin{equation}
H[P] = - \displaystyle \sum_{c}\mu_{c} \log \left[ \frac{\mu_{c}}{\epsilon^{d}} \right] = d \log \epsilon-D \log \epsilon+\mathrm{const}
\end{equation}
where $D$ is the information dimension of the integrated (natural) measure $\mu$, defined to be 
\begin{equation}
D = \underset{\epsilon \rightarrow 0^{+} }{\lim} \frac{ \sum_{c}\mu_{c} \log \mu_{c} } {\log \epsilon} .
\end{equation}
Applying this to the PMI through eq. (\ref{PMIEntropy}) then leads to
\begin{equation}
I(\tau)=I_{0}(\tau)-(D_{-}+D_{+}-D_{-+}) \log \epsilon
\end{equation}
where $D_{-+}$ is the information dimension of the joint distribution
of past and future, $D_{-}$ and $D_{+}$ the information dimensions
of the respective marginal distributions, and $I_{0}(\tau)$ the extrapolated
resolution-free PMI (note the dimensions of the underlying spaces cancel from $I(\tau)$ because  $ d_{-+}=d_{-}+d_{+}$).

Applying the equivalent analysis to the k-NN method we have
to be careful to insist that eq. (\ref{PMI}) is used, meaning in particular
that it is a neighbourhood of $k$ neighbours in the \emph{joint}
distribution which is taken to determine the ratio of joint and marginal
probability densities within the logarithm. With this understanding,
$\log \epsilon$ in the above expression for PMI can be written in terms of probability distribution $\frac{1}{D_{-+}} \log \left(k/N\right)$
leading to
\begin{equation}
I(\tau)=I_{0}(\tau)+\Gamma \log \left(N/k\right)
\end{equation}
where 
\begin{equation}
\Gamma=\frac{D_{-}+D_{+}-D_{-+}}{D_{-+}}
\end{equation}
is the relative information codimension.

Our first check is the accumulation point of period doubling of the logistic map, at which
the dynamics causally orbits its Cantor set attractor. In this case
all the information dimensions above are each equal to the fractal
dimension of the Cantor set, leading to $\Gamma=1$ in agreement with
our earlier observations and interpretation based on resolution limited
period. Fig. \ref{figFP} shows the directly measured PMI with apparent $\Gamma \simeq 0.9$ in fair agreement, where sampling over the attractor set is achieved by using different times of measurement. A uniform distribution of measurement times approximates uniform sampling of the unique invariant probability measure of the attractor.

\begin{figure}[th] 
	\centerline{\psfig{file=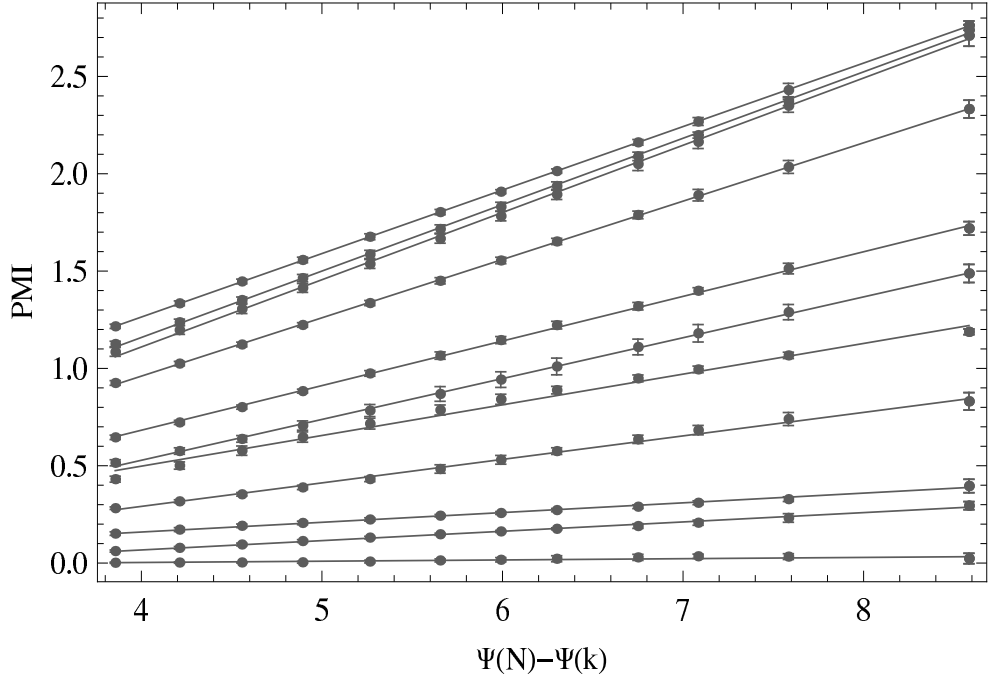,width=13cm, height = 9cm}}
	\vspace*{8pt}
	\caption{PMI for the standard map at 3000 iterations,  as a function of the probability resolution used to measure it, for map parameters $K=0.1,0.5,0.9,1.,1.1,1.2,1.5,2,3,4,6$ (top to bottom).  The resolution of probability is plotted as $\Psi(N)-\Psi(k) \simeq Ln(N/k)$, where $N=3000$ is the number of sample points used,  and $k$ is the rank of neighbour used in the measurement of Mutual Information (see text) which ranges from 1 to 64 across each plot. For each map parameter there is a clear linear dependence on this logarithmic resolution, consistent with fractal phenomenology and the interpretation of the slope as a relative information codimension.  The data have been averaged over five independent sets of such measurements with the error bars showing the $2\sigma$ error in each mean, and the drawn lines correspond to the slopes plotted in fig. \ref{fig4}. } \label{fig3}
	\end{figure}

The standard map provides a much more subtle test of this phenomenology.
This two-dimensional map
\begin{equation}
p'=\left(p+K\sin x\right)_{\mathrm{modulo}2\pi},\quad x'=\left(x+p'\right)_{\mathrm{modulo}2\pi}
\end{equation}
is strictly area-preserving in the $x,p$ plane (reduced modulo $2\pi$)
and has uniform invariant measure. Thus if we launch this dynamics
with random initial conditions, the marginal distributions (joint
of $x,p$ at fixed iteration) remain strictly uniform forever. The joint distribution between distant iterations of the standard map is far from simple and uniform, at least for moderate values of
the map parameter $K$. Fig. \ref{fig3}.  shows the measured PMI as a function
of the probability resolution $k/N$, displaying clear fractal behaviour for values of
$0 \leq K<6$.

	\begin{figure}[th] 
	\centerline{\psfig{file=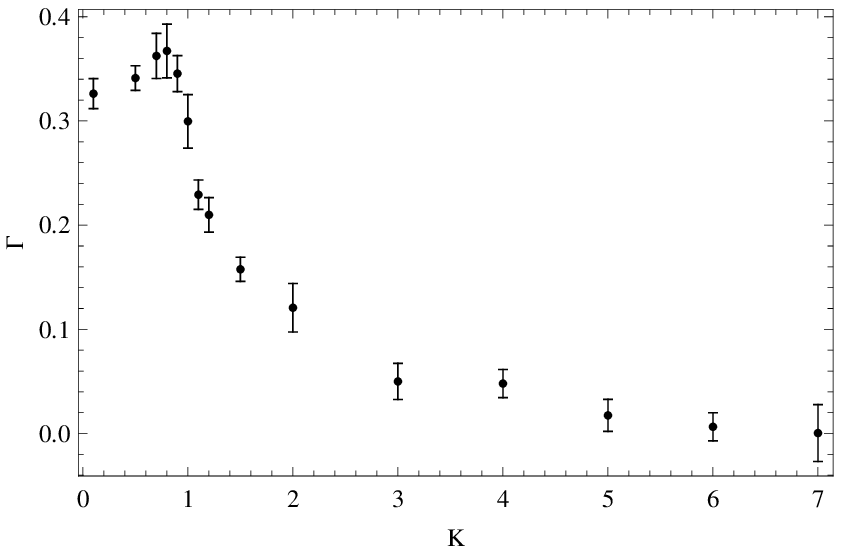,width=13 cm, height = 9cm}}
	\vspace*{8pt}
	\caption{Relative information codimension $\Gamma$ for the standard map (at 3000 iterations) as a function of map parameter $K$.  These are the slopes of the data shown in fig. \ref{fig3}, and the error bars are estimated as $2\sigma$  based on separate best fit to the five independent runs of data.  The intercept at $K=0$ matches theoretical expectation of $1/3$ (see text) and the fall to zero at large $K$  is consistent with dominance by chaotic dynamics. It is interesting that $\Gamma$ peaks in the vicinity of $K_g=0.97$ where the golden KAM curve breaks up, but anomalously slow dynamical relaxation in this region (see \cite{CS} and references therein) means that the peak may not reflect the limit of infinite iterations and hence true PPMI. } \label{fig4}
	\end{figure}

The corresponding estimates of the relative information codimension
are shown in fig. \ref{fig4}. For the left end point we can anticipate $\lim_{K\rightarrow0}\Gamma(K)=1/3$
on theoretical grounds, because this limit corresponds to a continuous
time Hamiltonian dynamics which is energy (Hamiltonian) conserving.
The joint distribution can therefore only explore $D_{-+}=4-1$ degrees
of freedom leading to $\Gamma=1/3$. Note this result depends on the
assumption that the shear in the dynamics between close energy shells
is sufficient to destroy correlation of their in-shell coordinates,
and we did correspondingly find that we had to use a large number of
map iterations for the expected behaviour to emerge.

The apparent peak in $\Gamma$ around $K=1$ is particularly interesting
because this is the vicinity of $K_{c}$ where momentum becomes diffusive,
closely associated (and often identified) with breakup of the golden
KAM curve at $K_{g}=0.971635..$ \cite{Greene,RSM85}. Dynamical anomalies have
been observed around this critical value of $K$ which might underlie
the peak we observe in PPMI. On the other hand corresponding long
time dynamical correlations pose a threat to whether our results are
adequately converged.

For larger $K$ our measurements are consistent with $\Gamma(K)=0$
for $K>K_{1}$ where $6<K_{1}<7$, and indeed the full PMI is within
uncertainty of zero in this regime, meaning the map appears fully chaotic to the level we can resolved.

\section{Conclusions}
We have shown that Persistent Mutual Information is a discriminating
diagnostic of hidden information in model dynamical systems, and the
Permanent PMI is a successful indicator of Strong Emergence.

The detailed behaviour of the logistic map is sufficiently re-entrant,
with periodicity and cascades of period multiplication intermingled
amidst chaos, that we are unlikely to have the last word on the full
quantitative behaviour of the PPMI as a function of map parameter
$\lambda$ beyond the first cascade. 

For the standard map, PPMI reveals some of the subtlety only otherwise
accessible through dynamical properties such as explicit orbits. Precise
relationships remain an open issue, particularly around critical map
parameter $K_{c}$. The observed fractal behaviour with a deficit
between the joint information dimension and those of the marginals
is we suggest a general phenomenology. Whether it reflects truly fractal
and multifractal behaviour in any particular case should rest on a
wider multifractal analysis of the joint probability measure, which
we intend to address in later work on a wider range of non-trivial
dynamical systems. 

Application to intrinsically stochastic systems and real world data
are outstanding challenges. We can however readily invoke a wide variety
of examples associated with ordering phenomena in statistical physics
where a dynamically persistent and spatially coherent order parameter
emerges. In these cases there is clearly PPMI in time, but also we
can consider just a time slice and let one spatial coordinate take
over the role of time in our PMI analysis. Spin Glasses are a key
instance where the two viewpoints are not equivalent: these have order
and hence Mutual Information persisting in time but not in space.

\section{Acknowledgements}
We acknowledge separate communications from P Grassberger, D Feldman and W Bialek which were particularly helpful in terms of references.
This research was supported by the EPSRC funding of Warwick Complexity Science Doctoral Training Centre, which fully funded MD, under grant EP/E501311/1.

	\bibliographystyle{ws-acs}
	\bibliography{bibliography}

		\end{document}